\documentclass[useAMS,usenatbib]{mnras} 

\usepackage[english]{babel}
\usepackage{journals}
\usepackage[pdftex]{graphicx,color} 
\usepackage[latin1]{inputenc}
\usepackage{graphics} 
\usepackage{amsfonts} 
\usepackage{amsmath}
\usepackage{multicol} 
\usepackage{layout} 
\usepackage{amssymb}
\usepackage{layouts}
\usepackage{multirow}
\usepackage{float}
\usepackage{tabularx}
\usepackage{siunitx}
\usepackage{subcaption}
\usepackage{bm}
\usepackage{verbatim}

\title[\textit{hybrid}-{\sc Lenstool}]{
\emph{hybrid}-{\sc Lenstool}: A self-consistent algorithm to model galaxy clusters with strong- and weak-lensing simultaneously 
}
\author[Niemiec et al.]
	{\parbox{\textwidth}{
	Anna Niemiec,$^{1}$\thanks{E-mail: \href{mailto:annaniem@umich.edu} {annaniem@umich.edu}} 
	Mathilde Jauzac,$^{2,3,4}$	
	Eric Jullo,$^{5}$ 
	Marceau Limousin,$^{5}$
	Keren Sharon,$^{1}$
	Jean-Paul Kneib,$^{6}$
	Priyamvada Natarajan,$^{7}$
	Johan Richard$^{8}$
	\\ \\ 
	}
	\vspace*{3pt}\\
$^{1}$Department of Astronomy, University of Michigan, 1085 South University Ave, Ann Arbor, MI 48109, USA\\
$^{2}$Centre for Extragalactic Astronomy, Department of Physics, Durham University, Durham DH1 3LE, UK\\
$^{3}$Institute for Computational Cosmology, Durham University, South Road, Durham DH1 3LE, UK\\
$^{4}$Astrophysics and Cosmology Research Unit, School of Mathematical Sciences, University of KwaZulu-Natal, Durban 4041, South Africa\\
$^{5}$Aix Marseille Univ, CNRS, CNES, LAM, 13388 Marseille, France \\
$^6$Laboratoire d'Astrophysique, Ecole Polytechnique F\'ed\'erale de Lausanne (EPFL), Observatoire de Sauverny, CH-1290 Versoix, Switzerland\\
$^7$Department of Astronomy, Yale University, New Haven CT 06511, USA\\
$^8$Univ Lyon, Univ Lyon1, Ens de Lyon, CNRS, Centre de Recherche Astrophysique de Lyon UMR5574, F-69230, Saint-Genis-Laval, France
   }
   
\begin{document}
\date{}
\maketitle
\label{firstpage}
\pagerange{\pageref{firstpage}--\pageref{lastpage}} \pubyear{2017}

\begin{abstract}
We present a new galaxy cluster lens modeling approach, \emph{hybrid}-\textsc{Lenstool}, that is implemented in the publicly available modeling software \textsc{Lenstool}. \emph{hybrid}-\textsc{Lenstool} 
combines a parametric approach to model the core of the cluster, and a non-parametric (free-form) approach to model the outskirts. 
\emph{hybrid}-\textsc{Lenstool} optimizes both strong- and weak-lensing constraints simultaneously (\emph{Joint-Fit}), providing a self-consistent reconstruction of the cluster mass distribution on all scales. In order to demonstrate the capabilities of the new algorithm, we tested it on a simulated cluster.
\emph{hybrid-}\textsc{Lenstool} yields more accurate reconstructed mass distributions than the former
 \emph{Sequential-Fit} approach where the parametric and the non-parametric models are optimized successively. Indeed, we show with the simulated cluster that the mass density profile reconstructed with a \emph{Sequential-Fit} deviates form the input by $2-3\sigma$ at all scales while the \emph{Joint-Fit} gives a profile that is within $1-1.5\sigma$ of the true value.
This gain in accuracy is consequential for recovering mass distributions exploiting cluster lensing and therefore for all applications of clusters as cosmological probes. 
Finally we found that the \textit{Joint-Fit} approach yields shallower slope of the inner density profile than the \emph{Sequential-Fit} approach, thus revealing possible biases in previous lensing studies.

\end{abstract}

\begin{keywords}
galaxies: clusters: general - gravitational lensing: strong - gravitational lensing: weak - cosmology: dark
\end{keywords}

\section{Introduction}
Gravitational lensing is the bending of the light emitted by a background source as it grazes past the gravitational potential of a foreground object called the lens \citep[for reviews see][]{kneib&natarajan2011,hoekstra2013,bartelmann2017}. The lens can be any type of objects with masses ranging from a planet to a massive galaxy cluster. Here, we focus on the lensing of a background galaxy population by a massive, foreground cluster of galaxies. Gravitational lensing is observed in two regimes defined by the intensity of the distortions created by the gravitational potential of the lens: the strong-lensing regime where background galaxies are highly distorted into gravitational arcs and multiple images; and the weak-lensing regime wherein the distortions of background galaxies are small and need to be treated statistically.

While gravitational lensing generated by a galaxy lens was first observed  in 1979 by \cite{walsh1979} with the multiply-imaged quasar Q0957\,+\,561A-B, astronomers had to wait until the late 1980's to confirm the first observation of a gravitational arc in a galaxy cluster \citep{lynds&petrosian1986,soucail1987,soucail1988}. Since then, gravitational lensing by galaxy clusters has emerged as a powerful tool to study the Universe. 
Indeed, gravitational lensing is a unique tool to map the mass distribution of the lenses as it is independent of their dynamical state, thereby providing crucial in-situ information on the physics of these objects. In the case of cluster lenses, a lot of work has been done from gravitational lensing mass maps and multi-wavelength analyses to constrain cluster physics \citep[e.g.,][]{kneib2003,natarajan2002a,clowe2004,bradac2006,merten2011,diego2015,eckert2015,jauzac2012,jauzac2015a,mahler2018,sharon2015,sharon2019}, and dark matter properties \citep[e.g.,][]{natarajan2002b,natarajan2017,bradac2008,harvey2015,harvey2016,harvey2017a,harvey2019,massey2015,massey2018,jauzac2016b,jauzac2018b}. Moreover, lensing can give us hints on galaxy evolution \citep[e.g.][]{natarajan1998,limousin2007,natarajan2009,leauthaud2012,leauthaud2015,li2015, sifon2015, niemiec2017}, and on the distant Universe as  lenses behave as cosmic telescopes and thus allow us to observe high-redshift galaxies  \citep[e.g.][]{atek2015,atek2018,alavi2016,bouwens2017,daloisio2014,ishigaki2018,kawamata2018}, study highly magnified galaxies at intermediate redshifts \citep[e.g. ][]{teplitz2000, bayliss2014, johnson2017, rigby2018, rivera-thorsen2019, bayliss2019, chisholm2019, sharon2019}, and/or lensed transients \citep[e.g.,][]{goobar2017,kelly2015,kelly2018,treu2016,jauzac2016a,diego2016,diego2018,diego2019,rodney2015,rodney2018,smith2018,smith2019}. Gravitational lensing can even be used to constrain cosmological parameters, as it is sensitive to the geometry of the Universe itself \citep{jullo2010,daloisio2011,caminha2016,acebron2017,suyu2018,wong2019,birrer2019}. 
However, to exploit gravitational lensing fully, the mass distribution of the lenses themselves requires to be modeled with high precision and accuracy.

There are currently two classes of lens modeling algorithms. The first one is based on \textit{parametric} mass models: the total mass distribution of the cluster is decomposed into a finite number of mass components divided into: \emph{(i)} the large-scale components, representing the mass contribution of the cluster dark matter halos and gas in the intra-cluster medium (ICM), and \emph{(ii)} the galaxy-scale components, representing the mass contribution of cluster galaxies. Dark matter halos are associated with individual cluster galaxies as smaller scale contributions to the overall mass distribution. The mass distribution of each component is then described by an analytical density profile, the most commonly employed ones being the Singular Isothermal Sphere potentials \citep[SIS, see e.g][]{binney&tremaine1987}, the Navarro-Frenk-White potentials \citep[NFW,][]{nfw1996}, and the Pseudo Isothermal Ellipsoidal Mass Distribution potentials \citep[PIEMD,][]{eliasdottir2007}. Current parametric mass modeling algorithms include \textsc{Lenstool} \citep{jullo2007}, \textsc{Glafic} \citep{oguri2010}, and \textsc{LTM} \citep{zitrin2012, zitrin2013}. 

The second class of algorithms relies on the so-called \textit{free-form} (or non-parametric) models. In this case, the cluster mass distribution is subdivided into a grid of mass ``pixels'', and the amplitude (and possibly the shape) of the pixels are optimized so that the overall mass distribution reproduces best the observed lensed image constraints. 
Free-form reconstruction algorithms include \textsc{SWUnited} \citep{bradac2005, bradac2009}, \textsc{WSLAP+} \citep{diego2005, diego2007, sendra2014, diego2016}, \textsc{Grale} \citep{liesenborgs2006, liesenborgs2009}, \textsc{LensPerfect} \citep{coe2008, coe2010}, \textsc{Lenstool} \citep[for weak-lensing mass reconstruction, see][]{jauzac2012, jullo2014}, SaWLens \citep{merten2009, merten2011}. We refer the reader to \citet[][]{kneib&natarajan2011} for a review on cluster lensing and mass modeling.

The two approaches appear to be complementary to model the different regions of the cluster: in the core, the sparse distribution of the strong-lensing constraints calls for a small number of free parameters, while the geometry of the constraints and the light distribution can give strong priors on the mass distribution, therefore favouring a parametric modeling approach. In the cluster outskirts, the density of constraints is high, and a more flexible free-form model with many mass ``pixels'' would allow a better tracing of the potentially irregular matter distribution, as well as the detection of the presence of (lower density than the cluster) infalling substructures \citep[see for example][for a comparison between the different types of models]{meneghetti2017,remolina-gonzalez2018}. Combining these approaches is the obvious next step, and with this in mind, we have developed a new version of the \textsc{Lenstool} software, \emph{hybrid}-\textsc{Lenstool}, that combines the parametric modeling approach in the cluster core, with a free-form grid model in the outskirts.

A key challenge when modeling galaxy clusters over an extended spatial scale is the nature of the different lensing constraints depending on the cluster region considered. In cluster cores, where the projected surface mass density is high, gravitational lensing is non-linear. This is the strong-lensing regime. Here background galaxies can be multiply-imaged in addition to being extremely distorted. In this case, the positions of the different multiple images of a same background galaxy are used to constrain the projected mass distribution of the lens. In the outskirts of clusters, the surface mass density is lower, images of background galaxies are thus only weakly distorted/sheared. There, gravitational lensing is on average mostly linear, this is the weak-lensing regime. A statistical approach is necessary to infer the projected mass distribution. The combination of the two types of constraints permits self-consistent modeling of the overall cluster mass density distribution.

In this paper, we present the new \emph{hybrid}-\textsc{Lenstool} algorithm that we developed in light of the large-scale weak lensing data that are collected via the Beyond Ultra-deep Frontier Fields And Legacy Observations (BUFFALO) survey \citep[GO-15117, PIs: Steinhardt \& Jauzac,][]{steinhardt2020}. With the availability of these high resolution \emph{Hubble Space Telescope} (\emph{HST}) observations of the outskirts of clusters, we developed a self-consistent model for the mass distribution of galaxy clusters that includes all scales, using a combination of both strong- and weak-lensing constraints.
The goal of the BUFFALO survey is to extend the \emph{HST} coverage of the 6 Hubble Frontier Field \citep[HFF,][]{lotz2018} clusters up to $\sim 3/4\times$R$_{\rm{vir}}$, thus allowing us to extend the HFF high resolution lens models beyond the cluster cores. 
Furthermore, we note that the current most up-to-date \textsc{Lenstool} strong-lensing model for Abell\,370, the first cluster fully observed in BUFFALO, requires a relatively strong external shear component in order to minimize the $\chi^{2}$ as described in \citet{lagattuta2019}. This strongly motivates the need for improvement of the lens modeling algorithm beyond the parametric version of \textsc{Lenstool}. Here, we present this new algorithm, \emph{hybrid}-\textsc{Lenstool}, and test the mass reconstruction with a simulated cluster, similar to the BUFFALO clusters. 

In Niemiec et al.\ (in prep.), we apply our newly formulated \emph{hybrid}-\textsc{Lenstool} to the extended observations of Abell\,370 obtained as part of BUFFALO, and show that the simultaneous modeling of both strong and weak lensing allows us to reduce the external shear component necessary in the parametric strong-lensing only mass model. This demonstrates the power of such an algorithm, and the mandatory need to model clusters consistently incorporating data from all scales comprehensively.

The outline of this paper is as follows: in Sect.~\ref{sec:method} we describe the method used to develop \emph{hybrid}-\textsc{Lenstool}, combining the parametric model used in the cluster core and the grid in the outskirts, and how these two modelling frameworks are optimized with both strong- and weak-lensing constraints simultaneously. In Sect.~\ref{sec:model_simu}, we test the algorithm on a simulated cluster, and quantify the improvement in mass modeling compared to the previous versions. We summarize the results and conclude with a discussion in Sect.~\ref{sec:conclusion}. Throughout this paper, we use a flat $\Lambda$CDM cosmology with $\Omega_{\rm{m}} = 0.27$ and $h_0 = 0.7$.

\section{Method}
\label{sec:method}

	\subsection{Lensing formalism}
	
Gravitational lensing formalism is based on the lens equation:
\begin{equation}
	\bm{\beta} = \bm{\theta} - \nabla{\psi}(\bm{\theta})
	\label{equ:lens_equ}
\end{equation}	
which relates the angular positions of the image and the source, $\bm{\theta}$ and $\bm{\beta}$ respectively, through the gradient of the so-called lensing potential, $\psi$, computed at the image position. The lensing potential is defined as
\begin{equation}
	\psi(\theta) = \frac{2}{c^2}\frac{D_{\rm{L}}D_{\rm{LS}}}{D_{\rm{S}}}\phi(\theta)\ ,
\end{equation}
where $D_{\rm{L}}$, $D_{\rm{S}}$, and $D_{\rm{LS}}$ represent the observer-lens, observer-source and lens-source angular diameter distances respectively, and $\phi(\theta)$ is the projected Newtonian potential of the lens. The Laplacian of the lensing potential is an indicator of the strength of the lens, and it can be linked using the Poisson equation to the projected surface mass density of the lens, $\Sigma(\theta)$, as:
\begin{equation}
	\Delta \psi (\theta) = 2\frac{\Sigma(\theta)}{\Sigma_{\rm{crit}}}\ ,
	\label{equ:poisson}
\end{equation}
where the critical surface density of the Universe, $\Sigma_{\rm{crit}}$, is defined as:
\begin{equation}
	\Sigma_{\rm{crit}} = \frac{c^2}{4\pi G}\frac{D_{\rm{S}}}{D_{\rm{L}}D_{\rm{LS}}}\ .
\end{equation}
The strength of a lens can thus be determined by comparing its surface mass density with the value of the critical surface density, $\Sigma_{\rm{crit}}$, at the corresponding source and lens redshifts. For a cluster at $D_{\rm L} = 1\, \rm{Gpc}$, and sources at $D_{\rm{S}} = 2\, \rm{Gpc}$ from the observer, the critical surface density is $\Sigma_{\rm{crit}} \sim 0.3\, \rm{g/cm}^2$. For a typical cluster, the core extends up to $0.5\,\rm{Mpc}$ with a central mass density $\rho_0 \sim 2\times10^{-25}\,\rm{g/cm}^3$ \citep{bahcall1977}. That gives a surface mass density, $\Sigma_{\rm{cluster}} \sim 0.3\,\rm{g/cm}^2 \sim \Sigma_{\rm{crit}}$. This quick estimation shows that the centre of clusters, where the density is the highest, can present over-critical strong-lensing regions where multiple images and gravitational arcs can be produced. In this regime, the observed position of multiple images relates to the source position through the lens equation (eq.\,\ref{equ:lens_equ}), which is degenerated in this case as for a given source position, $\beta$, multiple solutions, $\theta$, can exist.

In the regions of the cluster where the density is lower, $\Sigma_{\rm{cluster}} \ll \Sigma_{\rm{crit}}$, i.e. in the weak-lensing regime, both the shape distortion and magnification of the source are very small. These very weak distortions in the sub-critical regions require a statistical approach to measure the lensing signal.  The mapping from the unlensed to the lensed coordinates can be described by the Jacobian matrix, $\mathcal{A}$, also called the amplification matrix:
\begin{equation}
\mathcal{A}_{ij} = \frac{\partial \beta_i}{\partial \theta_j}\ .
\end{equation}
This matrix can be rewritten as:
\begin{equation}
\mathcal{A} = 
\begin{bmatrix}
 1 - \kappa	& 0			\\
 0			& 1 - \kappa	\\
\end{bmatrix}
+
\begin{bmatrix}
-\gamma_1	& -\gamma_2	\\
-\gamma_2	& \gamma_1	\\
\end{bmatrix}\ ,
\end{equation}
where we introduce the convergence, $\kappa$, that describes the magnification of the images, and the complex shear, $\gamma = \gamma_1 + i\gamma_2$, that describes the stretching of the images.
These parameters are derived from the lensing potential, $\psi(\theta)$ \citep{kneib&natarajan2011}:
\begin{equation}
	\kappa(\theta) =  \frac{1}{2}(\partial_1^2 + \partial_2^2)\psi(\theta)
\end{equation}
and
\begin{align}
\begin{aligned}
	\gamma_1(\theta) &= \frac{1}{2}(\partial_1^2 - \partial_2^2)\psi(\theta) \\
	\gamma_2(\theta) &= \partial_1\partial_2 \psi(\theta)
\end{aligned}
\end{align}
where $\partial_i$ represents the partial derivative with respect to $\theta_i$. We note that Equation \ref{equ:poisson} shows that the convergence is related to the projected surface mass density as $\kappa(\theta) = \Sigma(\theta)/\Sigma_{\rm{crit}}$.
	
\subsection{Parametric modeling of the cluster core}
	
In the central regions of galaxy clusters, i.e. in the strong-lensing regime, the geometry of multiple image systems and the distribution of cluster galaxies provide information on the priors for the matter distribution. It is therefore more appropriate to use so-called parametric models, which are described by physical quantities that allow a direct interpretation of the results. 
As described earlier, the matter distribution in this regime with this approach is then typically decomposed into cluster-scale and galaxy-scale haloes \cite{natarajan1997}.
The way these halos are modeled in \textsc{Lenstool} is described in detail in \citet{jullo2007}. Briefly, each halo is parametrized by its position in the sky $(x,y)$, projected ellipticity, $e$, and angle position, $\theta$. A number of parametric profiles are available to describe the distribution of dark matter within each halo, such as PIEMD \citep[][]{eliasdottir2007}, NFW \citep[][]{nfw1996}, or SIS. Each profile has a different set of parameters to describe the density slope.

Since the number of strong-lensing constraints is small compared to the number of galaxy-scale halos, the radial profile of each single galaxy-scale subhalos cannot be individually constrained.
Therefore, to decrease the number of free parameters, the mass of each subhalo is coupled to the luminosity of the galaxy it hosts using a global parametric mass-to-light relation. In practice, as initially proposed by \cite{natarajan1997}, subhalos are described with PIEMD profiles, in which free parameters are the core radius, $r_{core}$, the cut-off radius, $r_{\rm{cut}}$, and the velocity dispersion $\sigma_0$. These parameters are in turn related to the galaxy luminosity, $L$:
\begin{equation}
\begin{cases}
	\sigma_0 = \sigma_0^{\star}\left(\frac{L}{L^{\star}}\right)^{1/4},\\
	r_{\rm{core}} = r^{\star}_{\rm{core}}\left(\frac{L}{L^{\star}}\right)^{1/2},\\
	r_{\rm{cut}} = r^{\star}_{\rm{cut}}\left(\frac{L}{L^{\star}}\right)^{\alpha}
\end{cases}
\end{equation}
where $L^{\star}$ is the typical luminosity of a galaxy at the cluster redshift, and $r^{\star}_{\rm{cut}}$, $r^{\star}_{\rm{core}}$, and $\sigma_0^{\star}$ are its PIEMD parameters. These are the free parameters used to describe the mass of galaxy-scale subhalos. The total mass of one subhalo can then be written as:
\begin{equation}
	M = (\pi/G)(\sigma_0^{\star})^2r_{\rm{cut}}^{\star}(L/L^{\star})^{1/2+\alpha}.
\end{equation}
We fix $\alpha = 1/2$, following e.g \citet{jullo2009,richard2010}.
In the rest of this paper we denote by $\bm{\Theta}$ the vector containing the set of free parameters corresponding to the parametric part of the model.

\subsection{Grid modeling}

	In the cluster outskirts a more flexible approach is necessary to account for the potentially irregular shape of the cluster, and to allow for effective substructure detection. This can be best achieved using a non-parametric  model where the mass distribution is reconstructed using a grid of  ``mass  pixels''. As described in \citet{jullo2009, jullo2014}, the projected density (or convergence) field is decomposed into a grid of Radial Basis Functions (RBFs). More precisely, a grid covering the field to be modeled is set up, and a RBF is then fixed at each node, described by a truncated isothermal mass distribution (TIMD, the circularized version of a PIEMD).

The true convergence field, $\kappa(\theta)$, is therefore approximated as:
\begin{equation}
	\kappa(\theta) = \frac{1}{\Sigma_{\rm{crit}}} \sum_i v_i^2 f(||\theta_i - \theta||, s_i, t_i)\ ,
\end{equation}
where the RBF on grid node $\theta_i$ is defined as:
\begin{equation}
	f(R, s, t) = \frac{1}{2G}\frac{t}{t-s}\left( \frac{1}{\sqrt{s^2 + R^2}} - \frac{1}{\sqrt{t^2 + R^2}} \right)\ .
\end{equation}
For a TIMD profile, the weight of the RBF, $v^2$, is the velocity dispersion at the centre of the gravitational potential\footnote{The velocity distribution is usually noted $\sigma_0^2$, but we choose the notation $v^2$ to avoid confusion with the variance $\sigma$ that appears later in the text.}, and the RBF parameters, $s$ and $t$, represent the core and cut radii of the profile respectively. The core radius, $s$, is typically fixed to the distance between two nodes of the grid, and the cut radius to $s = 3t$.

The shear field can now be approximated by the RBFs as:
\begin{align}
\begin{aligned}
	\gamma_1(\theta) &= \sum_i v^2_i \Gamma_1(||\theta_i - \theta||, s_i, t_i) \\
	\gamma_2(\theta) &= \sum_i v^2_i \Gamma_2(||\theta_i - \theta||, s_i, t_i)\ ,
	\label{equ:shear_rbf}
\end{aligned}
\end{align}
where analytical expressions for $\Gamma_1$ and $\Gamma_2$ can also be derived \citep[see equation A8 in][]{eliasdottir2007}.

We denote by $\bm{w}$ the vector containing all the RBF weights, $v_i^2$, which corresponds to the set of free parameters for the grid model.
We set as a prior that the RBF weights, $v_i^2$, are positive, and following \citet{jullo2014}, that they are also described by a Poisson Probability Distribution Function (PDF):
\begin{equation}
    \text{P}(v_i^2) = \exp{(-v_i^2/q)}/q\ ,
\end{equation}
where $q$ is a nuisance parameter, described by the PDF:
\begin{equation}
    \text{P}(q) = q_0^2q\exp{(-q/q_0)},
\end{equation}
where the parameter $q_0$ is fixed to 10 following \citet{jullo2014}.

	\subsection{Likelihood definitions}
	
In the combined mass modeling, the two types of modeling frameworks described above (parametric + grid) are optimized jointly. To perform the optimization, two different types of constraints are used: the strong-lensing regions are constrained by the positions of the multiple images of a same background source, while the weak lensing regions are constrained by the shapes of the distorted images of the background sources producing the shear field. 

In order to consolidate these two approaches to produce a combined mass model, we take the following approach. The free parameters of the parametric part of the model (cluster-scale and galaxy-scale halos) are arranged in a vector $\bm{\Theta}$, and the grid model composed of $N$ RBFs with weight $v_i^2$, are ordered in a vector $\bm{w} = [v_1^2, ..., v_N^2]$. This allows us to combine and derive the total likelihood describing our model, which is written as:
\begin{equation}
	\mathcal{L}(\bm{\Theta}, \bm{w}) = \mathcal{L}_{\rm{SL}}(\bm{\Theta}, \bm{w}) \times \mathcal{L}_{\rm{WL}}(\bm{\Theta}, \bm{w})\ .
\end{equation}
We describe further in this section how we compute the strong- and weak-lensing likelihoods, $\mathcal{L}_{\rm{SL}}$, and $\mathcal{L}_{\rm{WL}}$ respectively.

		\subsubsection{Computing the strong lensing likelihood}

We consider a set of $M_{\rm{SL}}$ background sources strongly lensed so that each source $i$ has $n_i$ multiple images. Considering that the noise in the image position measurement of different images is uncorrelated, the noise covariance matrix is diagonal, and the likelihood can be written as: 
\begin{equation}
	\mathcal{L}_{\rm{SL}} = \prod_{i=1}^{M_{\rm{SL}}} \frac{1}{\prod_j \sigma_{ij}\sqrt{2\pi}}exp^{-\frac{\chi_i^2}{2}} ,
\end{equation}
where $\sigma_{ij}$ is the error on the position of image $j$ of the source $i$. The contribution of a multiple image system $i$ to the total $\chi^2$ can then be expressed as:
\begin{equation}
	\chi^2_i = \sum_{j=1}^{n_i} \frac{||x_{\rm{obs}}^j - x^j(\bm{\Theta}, \bm{w})||^2}{\sigma^2_{ij}}\ ,
\end{equation}
where $x^j_{\rm{obs}}$ is the measured position of the multiple image $j$, and $x^j(\bm{\Theta}, \bm{w})$ is the position of the image $j$ predicted by the model, in which the free parameters are $\bm{\Theta}$ for the parametric part, and $\bm{w}$ for the grid.

In the case of the combined model, we compute the $\chi^2$ in the source plane \citep[for a discussion on the pros and cons of computing the $\chi^2$ in the image or source plane, see][]{jullo2007}, which gives for one system:
\begin{equation}
	\chi^2_{S,i} = \sum_{j=1}^{n_i} \frac{||x^j_{\rm{S}}(\bm{\Theta}, \bm{w}) - < x_{\rm{S}}^j(\bm{\Theta}, \bm{w}) > ||^2}{\mu_j^{-2}\sigma^2_{ij}}\ ,
\end{equation}
where $x^j_{\rm{S}}(\bm{\Theta}, \bm{w})$ is the position of the source galaxy corresponding to the image $j$ projected to the source plane by the lens equation,
$< x_{\rm{S}}^j(\bm{\Theta}; \bm{w}) >$ is the barycenter of the positions of the source corresponding to all the images in system $i$, and $\mu_j$ is the magnification for image $j$. The source position $x^j_{\rm{S}}(\bm{\Theta}, \bm{w})$ can be calculated from the measured position of the image, $x^j_{\rm{obs}}$, by linearly adding the deflection angle of the parametric model and the RBFs, as:
\begin{equation}
    x^j_{\rm{S}}(\bm{\Theta}, \bm{w}) = x^j_{\rm{obs}} - \alpha(x^j_{\rm obs}, \bm{\Theta}) - \sum_i v^2_i \mathrm{A} (||x^j_{\rm obs} - x_i||, s_i, t_i)\ ,
\end{equation}
where $\alpha(x^j_{\rm obs}, \Theta)$ is the deflection angle produced at the observed image position by the mass distribution included in the parametric model, and $v^2_i \mathrm{A}(||x^j_{\rm obs} - x_i||, s_i, t_i)$ is the deflection angle produced at the image location by the RBF located at position $x_i$ \citep[see][for an analytical expression of A$(r,s,t)$]{eliasdottir2007}.

		\subsubsection{Computing the weak lensing likelihood}
		
We now consider a set of $\lambda$ background galaxies, each with a measured ellipticity, $e^i = [e_1^i, e_2^i]$, ordered in a vector of size $2\lambda$, $\bm{e} = [\bm{e_1}, \bm{e_2}]^{\top} = [e_1^1, ..., e_1^{\lambda},e_2^1, ..., e_2^{\lambda}]^{\top}$. We denote $\gamma^i = [\gamma_1^i, \gamma_2^i]$, the shear produced on the image of the galaxy $i$ by the grid model. The full shear vector of size $2\lambda$, $\bm{\gamma} = [\bm{\gamma_1}, \bm{\gamma_2}]^{\top} = [\gamma_1^1, ..., \gamma_1^{\lambda}, \gamma_2^1, ..., \gamma_2^{\lambda}]$, computed on the locations of the $\lambda$ background galaxies is then:
\begin{equation}
\bm{\gamma} = M_{\gamma w}\bm{w}\ ,
\end{equation}
where $\bm{w}$ are the weights of the $N$ RBFs, and $M_{\gamma w} = [\Delta_1, \Delta_2]^{\top}$ is a $2\lambda \times N$ matrix. Its elements represent the contributions of each unweighted RBF $j$ to the shear of image $i$:
\begin{align}
\begin{aligned}
	\Delta_1^{(i,j)} &= \frac{D_{\rm{LS}i}}{D_{\rm{OS}i}}\Gamma_1^i(||\theta_j - \theta_i||, s_j, t_j), \\
	\Delta_2^{(i,j)} &= \frac{D_{\rm{LS}i}}{D_{\rm{OS}i}}\Gamma_2^i(||\theta_j - \theta_i||, s_j, t_j)\ .
\end{aligned}
\end{align}

In the case of the combined model, part of the lens mass distribution is described by the parametric model, and the contribution to the total shear of this mass component must also be taken into account. We denote this shear component with $\bm{\gamma'}(\bm{\Theta})$, and we remind the reader that $\bm{\Theta}$ comprises the free parameters of the parametric part of the model.

In the linear weak-lensing approximation:
\begin{equation}
\bm{e} = 2\bm{\gamma} + 2\bm{\gamma'} + \bm{n}\ ,
\end{equation}
where $\bm{n}$ represents the intrinsic shapes of the galaxies in the source plane, described by a gaussian distribution with mean 0 and standard deviation $\sigma_n$. We note that the factor 2 comes from the complex ellipticity definition used in \textsc{Lenstool}, where its amplitude is expressed using the axis ratio, $r$, as $|e| = (1 - r^2)(1 + r^2)^{-1}$ \citep[corresponding to the $\chi$ notation in ][]{bartelmann2001}.

As the intrinsic galaxy ellipticity distribution is considered to be well described by a gaussian, 
the weak-lensing likelihood can be expressed as:
\begin{equation}
	\mathcal{L}_{\rm{WL}} = \frac{1}{Z_L}\exp^{-\frac{\chi^2_{\rm{WL}}}{2}}\ ,
\end{equation}
where 
\begin{equation}
	\chi^2_{\rm{WL}} = (\bm{e} - 2M_{\gamma w}\bm{w} - 2\bm{\gamma'(\Theta)})^{\top}N_{ee}^{-1}(\bm{e} - 2M_{\gamma w}\bm{w} - 2\bm{\gamma'(\Theta)})\ ,
\end{equation}
where $N_{ee} = <\bm{ee^{\top}}>$ is the covariance matrix of the measured ellipticities. Following \citet{jullo2014}, the matrix is considered diagonal, and its diagonal elements are expressed as:
\begin{equation}
	N_{ee}^{(i,i)} = \sigma^2_{\rm{m}} + \sigma^2_{\rm{int}}\ ,
\end{equation}
where $\sigma_{\rm{m}}$ is the measurement uncertainty, and $\sigma_{\rm{int}}$, the galaxy shape noise, which is defined as the scatter in the galaxy intrinsic shape distribution.

The normalization factor is written as $Z_L = \sqrt{(2\pi)^{2\lambda}\det N_{ee}}$.

\subsection{Implementing the MCMC sampling}

The PDFs of the lens model free parameters are sampled using the Monte-Carlo-Markov-Chain (MCMC) algorithm \textsc{BayeSys} \citep{skilling2004} implemented in \textsc{Lenstool}, as described in detail in \citet{jullo2007}.
In short, the free parameters  are sampled with 10 Markov Chains that explore the parameter space following a variant of the Metropolis-Hastings algorithm \citep{metropolis1953, hastings1970}. The selective annealing variant used in \textsc{BayeSys} ensures a progressive convergence of the chains from the prior to the posterior distribution without being trapped in any local minima.

As described in \cite{jauzac2012} and \cite{jullo2014}, \emph{hybrid}-{\sc Lenstool} also includes the \textsc{BayeSys} extension \textsc{MassInf}. This extension is useful when some of the free parameters have a linear contribution to the mass model. It allows us to find their values at each step of the MCMC through a Gibbs sampling, thus drastically decreasing the convergence time compared to a pure \textsc{BayeSys} sampling \citep[for details see][]{jauzac2012}. In our case, these linear parameters are the weights of the RBFs (see equation \ref{equ:shear_rbf}).

Throughout the sampling, \textsc{BayeSys+MassInf} does not assume that all the RBFs are necessary to reconstruct the mass distribution of the lens, but will rather use a number $n$ of them at each step. This effective number of RBFs is described with a geometric probability distribution:
\begin{equation}
    \text{P}(n) = (1 - c)c^{n - 1}\, \text{, with } c = \frac{\alpha}{\alpha + 1}\ .    
\end{equation}
Following \citet{jullo2014}, the parameter $\alpha$ is fixed at $2\%$ of the total number of RBFs.

We implement a block-wise sampling in the \textsc{BayeSys} algorithm, and sample the parametric model and the grid model parameters alternately.
In a regular Metropolis-Hasting sampling algorithm, at each step of the MCMC new values for all free parameters are drawn, and then accepted or rejected based on the values of the likelihood function. Given the large number of free parameters, it can become quite difficult, and onerous computing-wise, to find a new acceptable set of parameters. Instead, we here use component-wise sampling, i.e. we split the ensemble of parameters into two blocks, $\bm{\Theta}$ and $\bm{w}$, and update the blocks alternately. A more formal description of component-wise sampling  in general can be found for instance in \citet{johnson2009}.

\section{Tests of \emph{hybrid}-\textsc{Lenstool} on a simulated cluster}
\label{sec:model_simu}

	\subsection{The simulation}
	
We test \emph{hybrid}-\textsc{Lenstool} on a simulated cluster which replicates the mass distribution of the cluster Abell\,2744, as described in \citet{jauzac2016b}. This simulated cluster is composed of:	
\begin{itemize}
    \item two central large-scale potentials in the cluster core -- these potentials are modeled with PIEMDs, which parameters are summarized in Table\,\ref{tab:params}; the total mass of the PIEMDs is given by $M_{\rm{tot}} = 2\pi\rho_0\frac{r^2_{\rm{core}}r^2_{\rm{cut}}}{r_{\rm{core}} + r_{\rm{cut}}}$ \citep{eliasdottir2007}, where $\rho_0 = \frac{(1.46\sigma_0)^2}{2\pi G r_{\rm{core}}^2}$ \citep{jullo2007}. This gives $M_{\rm{C1}} = 1.03 \times 10^{13} M_{\sun}$, $M_{\rm{C2}} = 8.80 \times 10^{12} M_{\sun}$;
    \item six large-scale potentials to model surrounding substructures, located within 1 Mpc of the cluster centre. The substructures are also modeled with PIEMDs, with masses $M_{\rm{N}} = 2.41 \times 10^{12} M_{\sun}$, $M_{\rm{NW}} = 2.85 \times 10^{12} M_{\sun}$, $M_{\rm{S1}} = 7.82 \times 10^{11} M_{\sun}$, $M_{\rm{S2}} = 6.83 \times 10^{11} M_{\sun}$, $M_{\rm{S3}} = 1.99 \times 10^{12} M_{\sun}$ and $M_{\rm{S4}} = 8.98 \times 10^{11} M_{\sun}$;
    \item a catalogue of 246 galaxy-scale potentials, with parameters $r_{\rm{cut}}^* = 14\,\rm{kpc}$, $r_{\rm{core}}^* = 0.15\,\rm{kpc}$ and $\sigma_0^* = 155\,\rm{km/s}$. We note that positions and shapes of these potentials correspond to the true galaxy distribution in Abell\,2744 as measured in \citet{mahler2018}, and covers only the core region of the simulated cluster.
\end{itemize}

We use \textsc{Lenstool} to compute the deflection and shear maps corresponding to this mass distribution, and create the strong- and weak-lensing constraints by tracing back the positions and shapes of the background sources from the source to the image planes. We thus obtain:
\begin{itemize}
    \item 15 multiple image systems in the strong-lensing region covering the redshift range $1.5 < z < 5$, which roughly corresponds to the redshift range of multiple images with measured spectroscopic redshifts in A2744;
    \item a catalogue of weakly lensed background sources with a density of 45 sources/arcmin$^2$, and covering the redshift range $0.5 < z < 1.5$. This source density and redshift range correspond to the expected depth of the BUFFALO survey.  We draw the two components of the intrinsic ellipticities of the galaxies in a Gaussian PDF of width $\sigma_{\rm{int}} = 0.27$, while the size is considered constant. The sources are uniformly distributed in the source plane, and we remove the sources located in the strong-lensing region. To estimate the impact of the background source density, we perform a second mass reconstruction, with a weak-lensing source density $N_{\rm{s}} = 100$ sources/arcmin$^2$, which corresponds to a typical source density that can be obtained with deep HST observations \citep[see for example][]{jauzac2015a}. 
\end{itemize}
	
\begin{figure*} 
  \begin{center}
    \includegraphics[width=\textwidth]{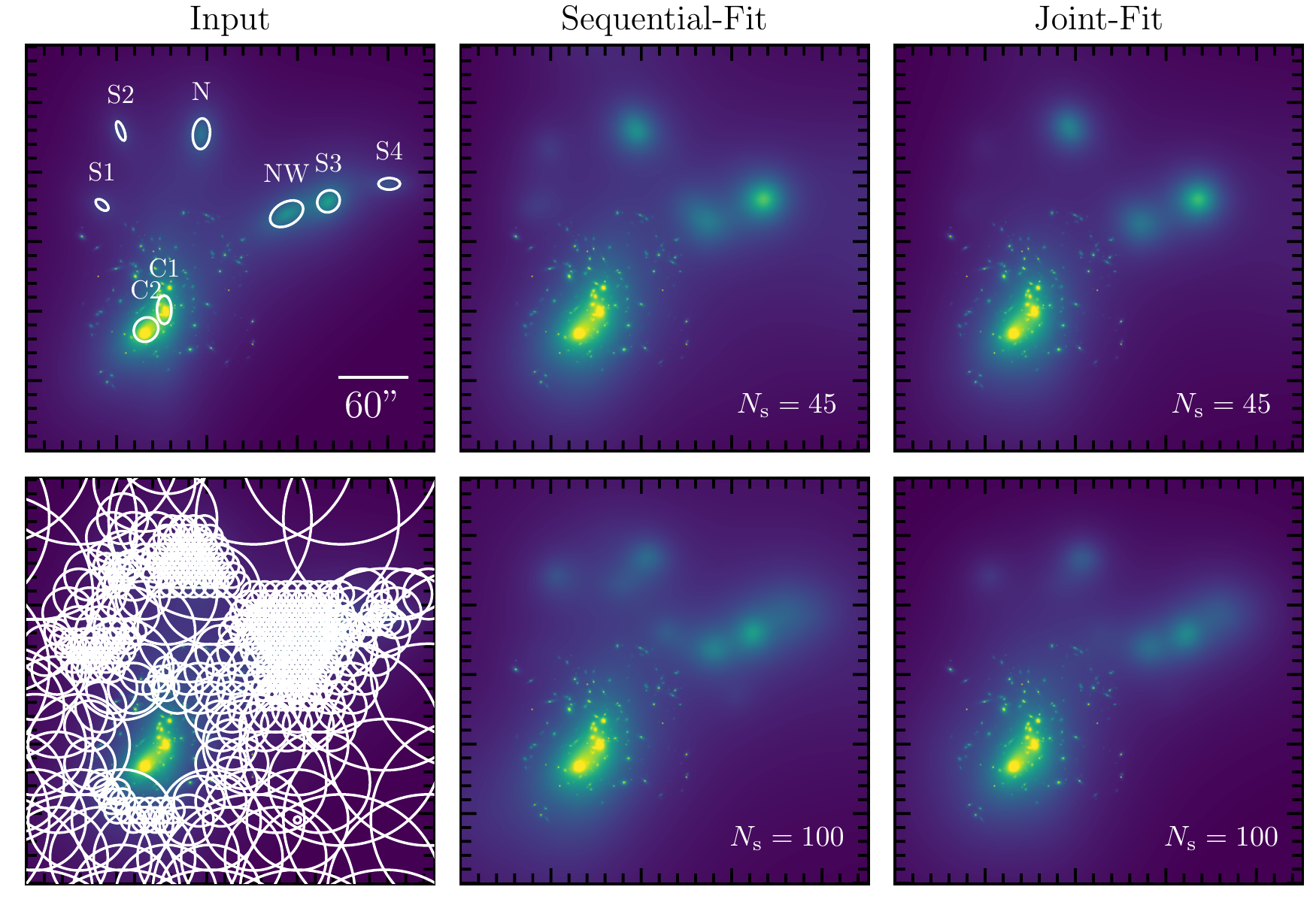}
    \caption{Projected mass maps. \textit{Left column:} Input simulation, with the large scale potentials shown as white ellipses (\textit{top panel}), and the potentials of the multiscale grid shown as white circles, which sizes are set to the core radii, $s$, of the potential (\textit{bottom panel}). \textit{Middle Column:} \emph{Sequential-Fit}, with source density $N_{\rm{s}} = 45$ sources/arcmin$^2$ (\textit{top}) and $N_{\rm{s}} = 100$ sources/arcmin$^2$ (\textit{bottom}). \textit{Right Column:} \emph{Joint-Fit}, with source density $N_{\rm{s}} = 45$ sources/arcmin$^2$ (\textit{top}) and $N_{\rm{s}} = 100$ sources/arcmin$^2$ (\textit{bottom}). The \emph{Sequential-} and \emph{Joint-Fit} mass maps are means over a 1000 MCMC samples.}
    \label{fig:input_mass}
  \end{center}
\end{figure*}

The projected mass distribution of the simulated cluster is presented in the left panel of Fig.\,\ref{fig:input_mass}, along with the large-scale potentials as white ellipses in the top panel, with  substructures named as in \citet{jauzac2016b}. Positions of the grid potentials are shown as white circles in the bottom panel, where the size of the circles is set to the potential core radii of these clumps.

\begin{table*}
	\begin{center}
		\begin{tabular}{|c|c c c c c c c c | c c | c c|}
		\hline
                    &   \multicolumn{8}{c|}{Input}                                  & \multicolumn{2}{c|}{Sequential-Fit}     & \multicolumn{2}{c|}{Joint-Fit}    \\		            
                    & C1    & C2    & N     & NW    & S1    & S2    &  S3   & S4    & C1                & C2                & C1                & C2                \\
		\hline
$x$ (arcsec)        & -2.1	& -17.7 & 29.8  & 103.0	& -55.5 & -39.4 & 139.0 & 191.1 & $-3.1 \pm 2.4$    & $-16.6 \pm 1.5$   & $-0.2 \pm 0.9$    & $-17.3 \pm 0.7$   \\
$y$ (arcsec)        & 1.4	& -15.7 & 153.3	& 84.3  & 91.9  & 155.7 & 95.0  & 110.1 & $1.8 \pm 2.9$     & $-15.4 \pm 1.0$   & $-0.1 \pm 1.1$    & $-15.7 \pm 0.6$   \\
$e$                 & 0.82	& 0.51	& 0.40	& 0.60  & 0.20  & 0.40  & 0.30	& 0.60  & $0.64 \pm 0.11$   & $0.38 \pm 0.08$   & $0.63 \pm 0.06$   & $0.41 \pm 0.05$   \\
$\theta$ ($\deg$)   & 90	& 45	& 85	& 30    & 140   & 110   & 40    & 0 	& $77 \pm 8$        & $39 \pm 9$        & $94 \pm 4$       `& $54 \pm 3$       \\
$r_{\rm core}$ (arcsec)& 18.8& 10.7  & 8.4	& 8.3   & 7.1   & 6.2   & 6.5  & 6.7	& $23.3 \pm 1.59$   & $13.8 \pm 2.5$    & $18.4 \pm 1.3$    & $12 \pm 1$       \\
$r_{\rm cut}$ (arcsec) & 221& 221   & 110   & 221   & 221   & 221   & 221   & 221   & $[221]$           & $[221]$           & $[221]$           & $[221]$           \\
$\sigma_0$ (km/s)   & 607	& 743   & 439	& 480   & 272   & 292   & 454	& 300   & $759 \pm 111$     & $754 \pm 102$     & $641 \pm 38$     & $762 \pm 38$     \\
\hline
		\end{tabular}
	\end{center}
\caption{ Parameters of the large-scale potentials for the simulated cluster used in this analysis. Potentials are modeled with PIEMDs: input (left), best-fit from the \emph{Sequential-Fit} (middle) and the \emph{Joint-Fit} (right).
The simulated cluster is located at redshift $z = 0.308$.}
\label{tab:params}
\end{table*}

\subsection{\textsc{Lenstool} mass reconstructions}

We perform two mass reconstructions of this simulated cluster: (1) the \emph{Sequential-Fit} -- we model the strong- and weak-lensing regions successively by first optimizing the parametric model in the cluster core with strong-lensing constraints; then we fix this part to its best-fit values, and optimize the grid model with weak-lensing constraints; (2) the \emph{Joint-Fit} -- we simultaneously optimize the parametric$+$grid models including both strong- and weak-lensing constraints, following the method presented in this paper.

We note that for both models presented in this section the parameters describing the galaxy-scale potentials are fixed to the input values. To test the implications of such choice, we perform a fit with these parameters set as free, and the \emph{Joint-Fit} does not show any improvement on the recovered values compared to the \emph{Sequential-Fit}.
	
	    \subsubsection{Sequential-Fit}
\paragraph*{Strong lensing modeling.}
We first reconstruct the cluster mass distribution following the method described in \citet{jauzac2015a}, i.e by modeling successively strong- and weak-lensing regions. Therefore we start by modeling the core of the cluster using strong-lensing constraints only, as if we had no knowledge on the presence of substructures in the outskirts. We reconstruct the mass distribution by optimizing the parameters describing the potentials C1 and C2. As is often done in similar analyses wherein the modeling is focused on the central region of the cluster, we keep the cut radius fixed to $r_{\rm{cut}} = 1000\,\rm{kpc}$. 

To be able to compare between the two methods we perform this strong-lensing optimization in the source plane. The best-fit values of the model parameters are presented in the middle part of Table\,\ref{tab:params}.

\paragraph*{Weak lensing component modeling.}

As a second step, we add a set of 353 RBFs to model the outskirts of the cluster. They are located at the nodes of a multi-scale grid. The resolution of the grid traces the mass distribution of the simulated cluster, with higher resolution in the densest region \citep[for more details on the multiscale grid, see for instance ][]{jullo2009, jauzac2012}. The resulting grid potentials have core radii varying between $s = 11\arcsec - 87\arcsec$ (i.e. $50 - 400\,\rm{kpc}$).
We remove the RBFs covering the cluster core, and model this central region with the best-fit values of the parametric model described in the previous paragraph
(see Table\,\ref{tab:params}). The grid of RBFs is created in a \textsc{Lenstool} input format using our set of publicly available scripts\footnote{\url{https://github.com/AnnaNiemiec/grid_lenstool}}.

We present the projected mass distribution maps resulting from these two successive models in the middle panel of Fig.\,\ref{fig:input_mass}, and refer to it as the \emph{Sequential-Fit}. The presented mass maps are the mean maps computed as the average of 1000 MCMC realizations.

\subsubsection{Joint-Fit}

We now reconstruct the cluster mass distribution by optimizing jointly the parametric and the grid models, with both strong- and weak-lensing constraints. We use the same priors on the free parameters as for the \emph{Sequential-Fit}, and the same multiscale grid. The best-fit parameters of the large-scale potentials are given in the right panel of Table\,\ref{tab:params}, and the resulting mass map is presented in the right panel of Fig\,\ref{fig:input_mass}. 
We refer to this mass reconstruction as the \emph{Joint-Fit}.

\subsection{Convergence diagnostics}
    
As we modified the MCMC sampling algorithm for \emph{hybrid}-\textsc{Lenstool}, we provide in this section a few diagnostics on the convergence of the chains. We use the 10 chains from the \emph{Joint-Fit} with $N_{\rm s} = 45$~sources/arcmin$^2$, where each chain was evolved for 100 steps after the burn-in phase.

We first verify the chains have converged by inspecting them visually. We then measure the serial correlation of the chains, and plot the auto-correlation functions (ACF) for each of them, i.e the amount of auto-correlation between the terms of the chain as a function of the lag, and verify that they rapidly decrease, and become consistent with 0 starting at lag $\sim 5$. Finally, we perform a Geweke diagnostic on each chain to check that they have reached a stationary state.

\subsection{Comparison of the \emph{Sequential-Fit} and the \emph{Joint-Fit} mass reconstructions}

\begin{figure} 
  \begin{center}
    \includegraphics[width = \linewidth]{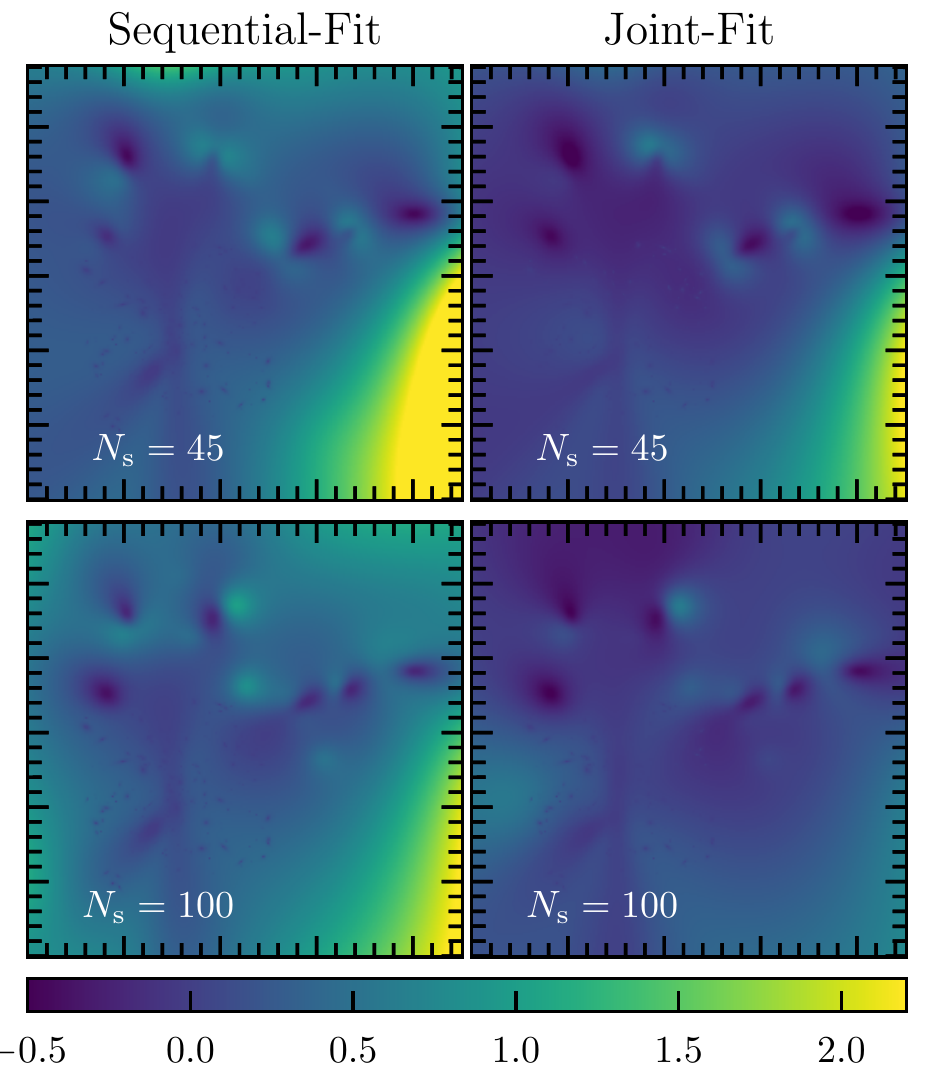}
    \caption{Normalized residual maps for the \emph{Sequential-Fit} (\text{left}) and \emph{Joint-Fit} (\textit{right}), and for source density $N_{\rm{s}} = 45$ sources/arcmin$^2$ (\textit{top}) and $N_{\rm{s}} = 100$ sources/arcmin$^2$ (\textit{bottom}).}
    \label{fig:mass_rapports}
  \end{center}
\end{figure}

To compare the reconstructed mass maps obtained with the different models, we compute the normalized residual maps, defined as $(M_{\rm{model}} - M_{\rm{input}})/M_{\rm{input}}$. We present the resulting maps on Fig.\,\ref{fig:mass_rapports} for the \emph{Sequential} (left panel) and \emph{Joint-Fit} (right panel), and for source densities $N_{\rm{s}} = 45$ sources/arcmin$^2$ (top), and $N_{\rm{s}} = 100$ sources/arcmin$^2$ (bottom).
The two models appear to well reproduce the simulated cluster, both in the core and the outskirts. As expected, mass reconstructions with a higher source density are closer to the input, i.e. they are better at detecting lower mass substructures such as S4 and S2, and trace better the true shape of substructures. For the two source densities, the \emph{Joint-Fit} mass reconstructions have a lower overall bias than the \emph{Sequential-Fit} reconstructions. 

Indeed, the best-fit parameters presented in Table\,\ref{tab:params} indicate that the combined modeling allows us to decrease the bias in the parametric model. When the core of the cluster is first modeled alone as in the \emph{Sequential-Fit}, the mass in the centre can often be overestimated to compensate for the mass missing in the outskirts. This is reflected in the best-fit values of the parameters $r_{\rm core}$ and/or $\sigma_{0}$ for both C1 and C2 clumps. This gives $M_{\rm{C1}} = (2.00 \pm 1.13)\times 10^{13} M_{\sun}$, and $M_{\rm{C2}} = (1.17 \pm 1.16)\times 10^{13} M_{\sun}$ for the \emph{Sequential-Fit}, and $M_{\rm{C1}} = (1.13 \pm 0.45)\times 10^{13} M_{\sun}$, and $M_{\rm{C2}} = (1.04 \pm 0.45)\times 10^{12} M_{\sun}$ for the \emph{Joint-Fit}.

\begin{figure} 
  \begin{center}
    \includegraphics{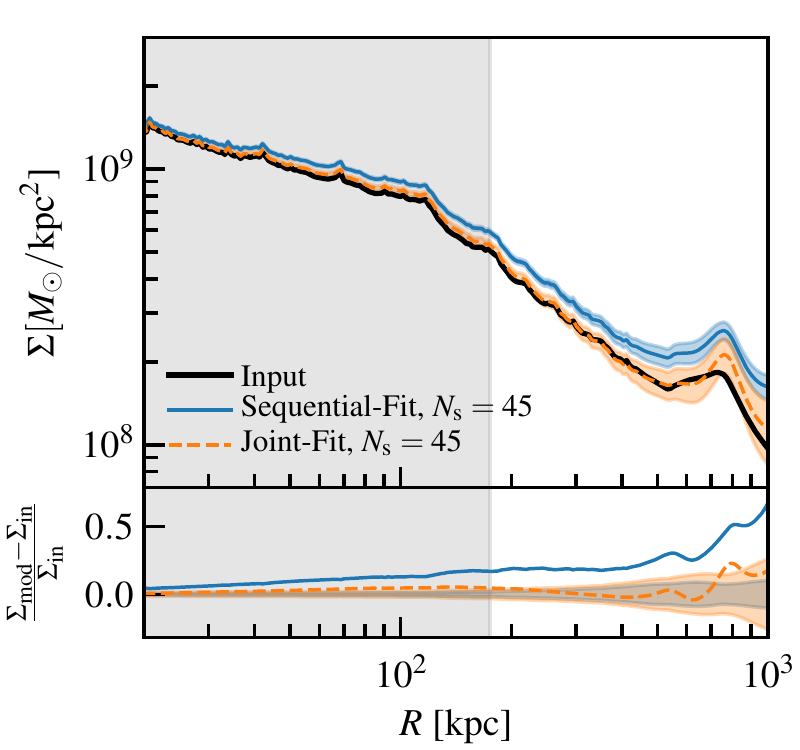}
    \caption{\emph{Top panel:} Projected mass density profiles of the input simulation (black line), the \emph{Sequential-Fit} (solid blue line) and the \emph{Joint-Fit} (dashed orange line) for the source density $N_{\rm{s}} = 45$ sources/arcmin$^2$. \emph{Bottom panel:} Relative deviation for the \emph{Sequential-Fit} (solid blue line) and \emph{Joint-Fit} (orange dashed line). The shaded regions represent the relative errors for both models. The radial range shown by the grey shaded area corresponds to the strong-lensing region of the simulated cluster.}
    \label{fig:mass_profiles}
  \end{center}
\end{figure}

\begin{figure} 
  \begin{center}
    \includegraphics{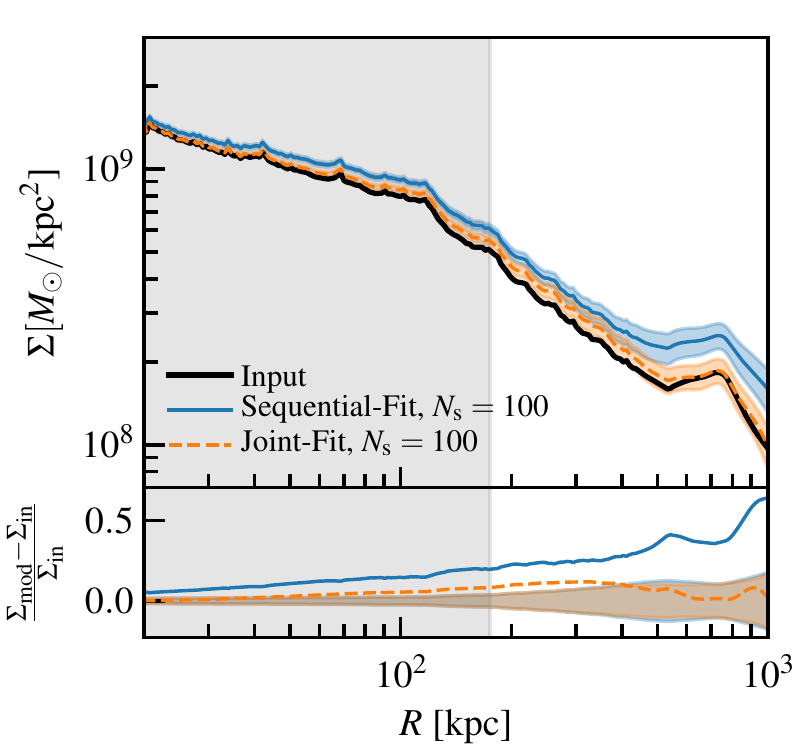}
    \caption{Same as Fig.\,\ref{fig:mass_profiles} for source density $N_{\rm{s}} = 100$ sources/arcmin$^2$.}
    \label{fig:mass_profiles_d2}
  \end{center}
\end{figure}

To quantify the model deviations from the input simulation, we measure the projected density profiles for each of the models. For each of the 1000 model realizations corresponding to the MCMC samples, we compute a radial density profile by azimuthally averaging the reconstructed mass maps, taking the cluster centre close to the centre of the mass clump, C1. We then average the 1000 profiles, and compute error bars by taking the standard deviation over all 1000 measurements. Figures\,\ref{fig:mass_profiles} and \ref{fig:mass_profiles_d2} show the density profiles for source densities $N = 45$, and 100 sources/arcmin$^2$ respectively. In both figures, the top panel shows the density profiles, while the bottom panel shows the relative deviations of models as lines and relative errors as shaded areas. 

For the two background source number densities, the \emph{Sequential-Fit} deviates from the simulation by $\sim 2-3 \sigma$ on all scales and consistently predicts a higher mass density. As explained before, this is due to the overestimation of the parameters which define the amplitude and size of the central clumps, as this impacts the mass distribution over all scales. Fitting the distribution in the core and in the outskirts self-consistently allows us to avoid this systematic effect. Indeed, Fig.\,\ref{fig:mass_profiles} and Fig.\,\ref{fig:mass_profiles_d2} show that the \emph{Joint-Fit} gives a density profile within $1 - 1.5\sigma$ of the true value over all scales.
Finally, the mass density profiles confirm that increasing the source number density allows a better recovery of the shapes of substructures, and produces a smoother mass distribution.

We also compute the integrated mass profiles, and derive the total mass within 1\,Mpc, for each model. The mass of the simulated cluster is $M(< 1\,\rm{Mpc}) = 9.02 \times 10^{14}M_{\sun}$. As can be inferred from the density profiles, the \emph{Sequential-Fit} overestimates the total enclosed mass, i.e. $M(< 1\,\rm{Mpc}) = (11.52 \pm 0.68) \times 10^{14}M_{\sun}$ ($(12.43 \pm 1.45) \times 10^{14}M_{\sun}$) for $N = 45$  (100) sources/arcmin$^2$. The \emph{Joint-Fit} shows a reduced bias on the total mass estimate, and gives values consistent with the true mass: $M(< 1\,\rm{Mpc}) = (9.43 \pm 0.92) \times 10^{14}M_{\sun}$ ($(9.70 \pm 0.82) \times 10^{14}M_{\sun}$) for $N = 45$  (100) sources/arcmin$^2$.

A commonly used indicator to quantify the goodness of fit for a strong-lensing model is the root-mean-square, noted as rms, i.e. the distance between the observed positions of multiple images and the ones predicted by the best fit model. We compute the best-fit rms values for the different models, and find that the combined modeling with $N_{\rm s} = 45$~sources/arcmin$^2$ indeed decreases the rms to 0.57\arcsec, compared to 0.63\arcsec\, for the strong-lensing only model of the cluster core. The rms value is further slightly decreased to 0.54\arcsec\, for the \emph{Joint-Fit} with $N_{\rm s} = 100$, which points that improving the modeling of substructures impacts the quality of fit in the cluster core, in agreement with conclusions from \citet{acebron2017} analysis. We note that the decrease in rms value can only be considered as a significant estimator of the goodness of fit when comparing between different models of the same cluster, as is the case with our analysis.

The last point we would like to address is the speed of the algorithm. The \emph{Joint-Fit} mass reconstruction performed with \emph{hybrid-}\textsc{Lenstool} is significantly slower than the one performed with the regular version of \textsc{Lenstool}. For instance, on a standard laptop, for the models presented here with the source density $N_{\rm s} = 45$, the \emph{Joint-Fit} can be obtained in 15\,mn, and the \emph{Sequential-Fit} in 5\,mn (3\,mn for the parametric model and 2\,mn for the grid). While this result could be expected and is not particularly problematic in this case, this time difference could become more of an issue when modeling very complex systems. In addition, keeping the parameters of the galaxy-scale potentials free further slows down the modeling, and in a larger proportion in the case of the \emph{Joint-Fit} compared to the \emph{Sequential-Fit}. Indeed at each step of the optimization, it requires to compute the shear produced by each of the cluster galaxies (which can be hundreds) at the position of each weakly lensed galaxies (which can be thousands). This can be made faster by implementing parallelisation in the computation of the weak-lensing likelihood, as already done for the strong-lensing one.

\section{Summary and conclusions}
\label{sec:conclusion}

In this paper, we present \emph{hybrid}-\textsc{Lenstool}, a new method implemented in the publicly available lens reconstruction algorithm \textsc{Lenstool}. It combines strong- and weak-lensing constraints to self-consistently reconstruct the cluster mass distribution at all scales. \emph{hybrid}-\textsc{Lenstool} combines a parametric model in the cluster core with a free-form grid model in the outskirts. It takes advantage of the complementary strengths of the two types of modeling to recover the shape and amplitude of the mass distribution with high precision. 

We tested this new method on a simulated cluster composed of a bi-modal mass distribution in the core and 6 massive substructures in the outskirts. We found that the \emph{Joint-Fit} modeling recovers well the shape and position of the substructures, and gives a more accurate reconstructed mass density profile for the cluster compared to a \emph{Sequential-Fit}, where the core and the outskirts of the cluster are modeled separately (method that was used in the past). In addition, the \emph{Joint-Fit} performs better at predicting the position of multiple images in the cluster core, reducing the rms from 0.63\arcsec\, to 0.57\arcsec.

After demonstrating the power of this new algorithm with simulated data in this method paper, we will present the first application of \emph{hybrid-}\textsc{Lenstool} to real observations in a forthcoming paper (Niemiec et al., \emph{in prep}). As mentioned before, the ongoing BUFFALO survey \citep[GO-15117, PIs: Steinhardt \& Jauzac,][]{steinhardt2020} is extending the HST coverage of the 6 HFF clusters \citep[PI: Lotz,][]{lotz2018}, and will complement the strong-lensing constraints in the core of these clusters with high resolution weak-lensing data. Combining these datasets, i.e. high resolution constraints both in the strong- and weak-lensing regions of the clusters, with \emph{hybrid-}\textsc{Lenstool} will produce high-precision models of the mass distribution of the HFF clusters up to $\sim 3/4\,R_{\rm{vir}}$. Higher fidelity mass distributions for cluster lenses are important for utilizing the full potential of clusters to probe dark matter properties and cluster physics, study the distant Universe that they magnify, and be used as cosmological probes, as observed lensing effects are not impacted by the dynamical complexity of clusters.

\section*{Acknowledgments}
AN would like to thank Jonathan Vacher for the helpful discussions. Support for BUFFALO surveys Program number GO-15117 was provided by NASA through a grant from the Space Telescope Science Institute, which is operated by the Association of Universities for Research in Astronomy, Incorporated, under NASA contract NAS5-26555. MJ is supported by the United Kingdom Research and Innovation (UKRI) Future Leaders Fellowship `Using Cosmic Beasts to uncover the Nature of Dark Matter' [grant number MR/S017216/1]. This project was also supported by the Science and Technology Facilities Council [grant number ST/L00075X/1]. AN, PN, KS acknowledge support from the BUFFALO program. ML, EJ, AN acknowledges CNRS and CNES for support.

\bibliographystyle{mn2e}
\bibliography{lensing.bib}

\label{lastpage}
\end{document}